\documentclass[conference]{IEEEtran}
\IEEEoverridecommandlockouts

\usepackage{cite}
\usepackage{graphicx}
\usepackage{amsmath}
\usepackage{mdwmath}
\usepackage{amssymb}
\usepackage{mdwtab}
\usepackage{stfloats}
\usepackage[footnotesize]{subfigure}
\usepackage{amsmath,amsthm}
\usepackage{threeparttable}
\usepackage{xcolor}
\usepackage{url}
\usepackage{algpseudocode}
\usepackage{multicol}
\usepackage{multirow}
\usepackage{setspace}
\usepackage{bm}
\usepackage{enumitem}
\usepackage{textcase}
\usepackage{balance}
\usepackage{comment}
\usepackage{tikz}
\usepackage{pgfplots}
\pgfplotsset{compat=newest}
\pgfplotsset{plot coordinates/math parser=false}
\newlength\figureheight
\newlength\figurewidth

\algrenewcommand{\algorithmicrequire}{\textbf{Input:}}
\algrenewcommand{\algorithmicensure}{\textbf{Output:}}

\newtheorem{remark}{\it  Remark}

\def\BibTeX{{\rm B\kern-.05em{\sc i\kern-.025em b}\kern-.08em
    T\kern-.1667em\lower.7ex\hbox{E}\kern-.125emX}}

\newcommand*\circled[1]{\tikz[baseline=(char.base)]{
            \node[shape=circle,draw,inner sep=.5pt] (char) {#1};}}

\usepackage{hyperref}
\hypersetup{
    colorlinks=true,
    linkcolor=red,
    citecolor=red,
    urlcolor=blue,
    breaklinks=true}
\usepackage[nolist]{acronym}
\begin{acronym}[ACRONYM]
\acro{AI}{artificial intelligence}
\acro{AoA}{angle-of-arrival}
\acro{AoD}{angle-of-departure}
\acro{BS}{base station}
\acro{BP}{belief propagation}
\acro{CDF}{cumulative density function}
\acro{CFO}{carrier frequency offset}
\acro{CRB}{Cram\'er-Rao bound}
\acro{DA}{data association}
\acro{D-MIMO}{distributed multiple-input multiple-output}
\acro{DL}{downlink}
\acro{EM}{electromagnetic}
\acro{FIM}{Fisher information matrix}
\acro{GDOP}{geometric dilution of precision}
\acro{GNSS}{global navigation satellite system}
\acro{GPS}{global positioning system}
\acro{IP}{incidence point}
\acro{IQ}{in-phase and quadrature}
\acro{ISAC}{integrated sensing and communication}
\acro{ICI}{inter-carrier interference}
\acro{JCS}{Joint Communication and Sensing}
\acro{JRC}{joint radar and communication}
\acro{JRC2LS}{joint radar communication, computation, localization, and sensing}
\acro{IMU}{inertial measurement unit}
\acro{IOO}{indoor open office}
\acro{IoT}{Internet of Things}
\acro{IRN}{infrastructure reference node}
\acro{KPI}{key performance indicator}
\acro{LoS}{line-of-sight}
\acro{LS}{least-squares}
\acro{MCRB}{misspecified Cram\'er-Rao bound}
\acro{MIMO}{multiple-input multiple-output}
\acro{ML}{maximum likelihood}
\acro{mmWave}{millimeter-wave}
\acro{NLoS}{non-line-of-sight}
\acro{NR}{new radio}
\acro{OFDM}{orthogonal frequency-division multiplexing}
\acro{OTFS}{orthogonal time-frequency-space}
\acro{OEB}{orientation error bound}
\acro{PEB}{position error bound}
\acro{VEB}{velocity error bound}
\acro{PRS}{positioning reference signal}
\acro{QoS}{Quality of Service}
\acro{RAN}{radio access network}
\acro{RAT}{radio access technology}
\acro{RCS}{radar cross section}
\acro{RedCap}{reduced capacity}
\acro{RF}{radio frequency}
\acro{RIS}{reconfigurable intelligent surface}
\acro{RFS}{random finite set}
\acro{RMSE}{root mean squared error}
\acro{RTK}{real-time kinematic}
\acro{RTT}{round-trip-time}
\acro{SLAM}{simultaneous localization and mapping}
\acro{SLAT}{simultaneous localization and tracking}
\acro{SNR}{signal-to-noise ratio}
\acro{ToA}{time-of-arrival}
\acro{TDoA}{time-difference-of-arrival}
\acro{TR}{time-reversal}
\acro{TXRX}[TX/RX]{transmitter/receiver}
\acro{Tx}{transmitter}
\acro{Rx}{receiver}
\acro{UE}{user equipment}
\acro{UL}{uplink}
\acro{UWB}{ultra wideband}
\acro{XL-MIMO}{extra-large MIMO}

\acro{PL}{protection level}
\acro{TIR}{target integrity risk}
\acro{IR}{integrity risk}
\acro{RAIM}{receiver autonomous integrity monitoring}
\acro{1D}{one-dimensional}
\acro{3D}{three-dimensional}
\acro{SS}{solution separation}
\acro{PMF}{probability mass function}
\acro{PDF}{probability density function}
\acro{GM}{Gaussian mixture}
\acro{CCDF}{complementary cumulative distribution function}

\end{acronym}

\begin{document}

\title{Bayesian Integrity Monitoring for Cellular Positioning --- A Simplified Case Study
\thanks{This project has received funding from the European Union's Horizon 2020 research and innovation programme under grant agreement No. 101006664. The authors would like to thank all partners within Hi-Drive for their cooperation and valuable contribution. This work is also  supported in part by Spanish R+D Grant PID2020-118984GB-I00 and in part by the Catalan ICREA Academia Programme.
}}

\author{\IEEEauthorblockN{Liqin~Ding\IEEEauthorrefmark{1},  Gonzalo~Seco-Granados\IEEEauthorrefmark{2}, Hyowon~Kim\IEEEauthorrefmark{1},\\  Russ~Whiton\IEEEauthorrefmark{3}, Erik~G.~Ström\IEEEauthorrefmark{1}, Jonas~Sjöberg\IEEEauthorrefmark{1}, Henk~Wymeersch\IEEEauthorrefmark{1}}
	\IEEEauthorblockA{\IEEEauthorrefmark{1}Department of Electricl Engineering, Chalmers University of Technology, Gothenburg, Sweden\\
 \IEEEauthorrefmark{2}Department of Telecommunications and
Systems Engineering, Universitat Autonoma de Barcelona, Barcelona, Spain\\
	\IEEEauthorrefmark{3}Volvo Car Corporation, Gothenburg, Sweden
	}
}
	
\maketitle

\begin{abstract}
    Bayesian receiver autonomous integrity monitoring (RAIM) algorithms are developed for the snapshot cellular positioning problem in a simplified one-dimensional (1D) linear Gaussian setting. Position estimation, multi-fault detection and exclusion, and protection level (PL) computation are enabled by the efficient and exact computation of the position posterior probabilities via message passing along a factor graph. Computer simulations demonstrate the significant performance improvement of the proposed Bayesian RAIM algorithms over a baseline advanced RAIM algorithm, as it obtains tighter PLs that meet the target integrity risk (TIR) requirements.
\end{abstract}


\section{Introduction}
\label{sec:1}

The estimation of position and its associated confidence level is central to numerous engineering applications. One can intuitively understand the utility of receiving both a position estimate and a confidence estimate by the blue circle with changing radius around the ego position in smartphone navigation \cite{mckenzie2016assessing}. 
\textit{Integrity} assurance mechanisms and \ac{RAIM} algorithms have long been introduced into \acp{GNSS} \cite{walter2003integrity} to ensure a high degree of confidence in the information provided by the systems to the end user. Safety-critical applications in aviation \cite{blanch2015baseline}, rail \cite{marais2017survey}, and automotive \cite{Reid2019Localization} benefit greatly from strong {integrity} guarantees in positioning systems. The rigorous quantification of integrity is typically done through the formulation of an upper bound of instantaneous position error, termed \ac{PL}, to meet the required confidence level (i.e. the probability that the actual error is below \ac{PL}), often given in the form of one minus the so-called \ac{TIR}. It is expected that the computed \acp{PL} are tight enough so that their utility in other applications can be maximized \cite{larson2017conservatism}.

As new safety-critical applications emerge in urban and indoor scenarios, where \acp{GNSS} suffers from poor coverage, cellular positioning \cite{dwivedi2021positioning} promises to provide a reliable complementary solution, for which integrity support becomes crucial \cite{whiton2022cellular,maaref2021autonomous_raim}. In \cite{whiton2022cellular}, the importance of integrity for cellular localization is argued. Specific errors such as multipath biases and blockages are studied in \cite{maaref2021autonomous_raim} and mitigated using \ac{RAIM}.  
 Integrity support for \ac{GNSS} assistance has been incorporated into the latest 3GPP Release-17 standard \cite{3gpp.38.857}, and radio standalone positioning with integrity guarantee is expected from Release-18 onward \cite{3gpp.38.859}. This emphasizes the importance of RAIM methods for cellular positioning, providing integrity guarantees and tight \acp{PL} at the same time.

\ac{RAIM} methods can be grouped into \emph{traditional \ac{RAIM} algorithms} and \emph{Bayesian \ac{RAIM} methods}. Traditional \ac{RAIM} algorithms detect and exclude faulty measurements by performing rounds of consistency checks on the statistics (e.g. range residuals and position estimates) associated with the fault patterns (i.e. the alternative hypotheses in the language of statistical hypothesis testing) that would otherwise lead to large errors \cite{zhu2018gnss}. This frequentist approach relies on redundant measurements and does not directly lead to instantaneous position error probability distributions. To avoid underestimating tail risk, the formulation of solvable \ac{PL} equations has to conservatively overbound the error distribution. Consequently, the computed \acp{PL} tends to be loose. In contrast, Bayesian \ac{RAIM} methods \cite{pesonen2011framework, zhang2015new, gupta2019particle,  gabela2021case} aim to find the posterior probability distribution of position error directly based on the information contained in the measurement models (prior) and actual measurements (evidence). It can be expected that the computed \acp{PL} are tight, since, in theory, all the position information contained in the measurements can be preserved in the posterior. The downside of Bayesian \ac{RAIM} is its potentially high complexity associated with posterior computation, particularly when the number of unknown parameters is large and the problem model admits no closed-form expressions (which, unfortunately, is the general situation). Therefore, a major challenge of Bayesian methods is to find computationally efficient implementations. To date, research work on Bayesian \ac{RAIM} methods is still very limited, and most of them have adopted Monte Carlo algorithms, such as particle filters \cite{gupta2019particle,  gabela2021case} and Gibbs sampler \cite{zhang2015new}, for posterior distribution computation.

In this paper, we consider the \ac{RAIM} problem for snapshot\footnote{The connection of user position between epochs is ignored, although it can be included naturally through the prediction step in the proposed Bayesian (filter) framework, and we assume that all measurements are taken at the same time instance in one snapshot.} cellular positioning, providing three distinct contributions: (i) we propose a novel factor graph-based Bayesian RAIM method to compute position and \ac{PL},  
including multi-fault detection and exclusion; (ii) we evaluate the method in a simplified \ac{1D} linear Gaussian scenario and compare its performance with a baseline advanced RAIM algorithm \cite{blanch2015baseline} using Monte-Carlo simulation; (iii) we demonstrate that, while fulfilling the \ac{TIR} requirement, the resultant \acp{PL} are much tighter than the baseline algorithm due to the exact computation of the posterior probability density by the new method, thus greatly improving the availability of the system.

\section{Problem Formulation}
\label{sec:2}

\begin{figure}[!t]
	\centering
\includegraphics[width=.88\linewidth]{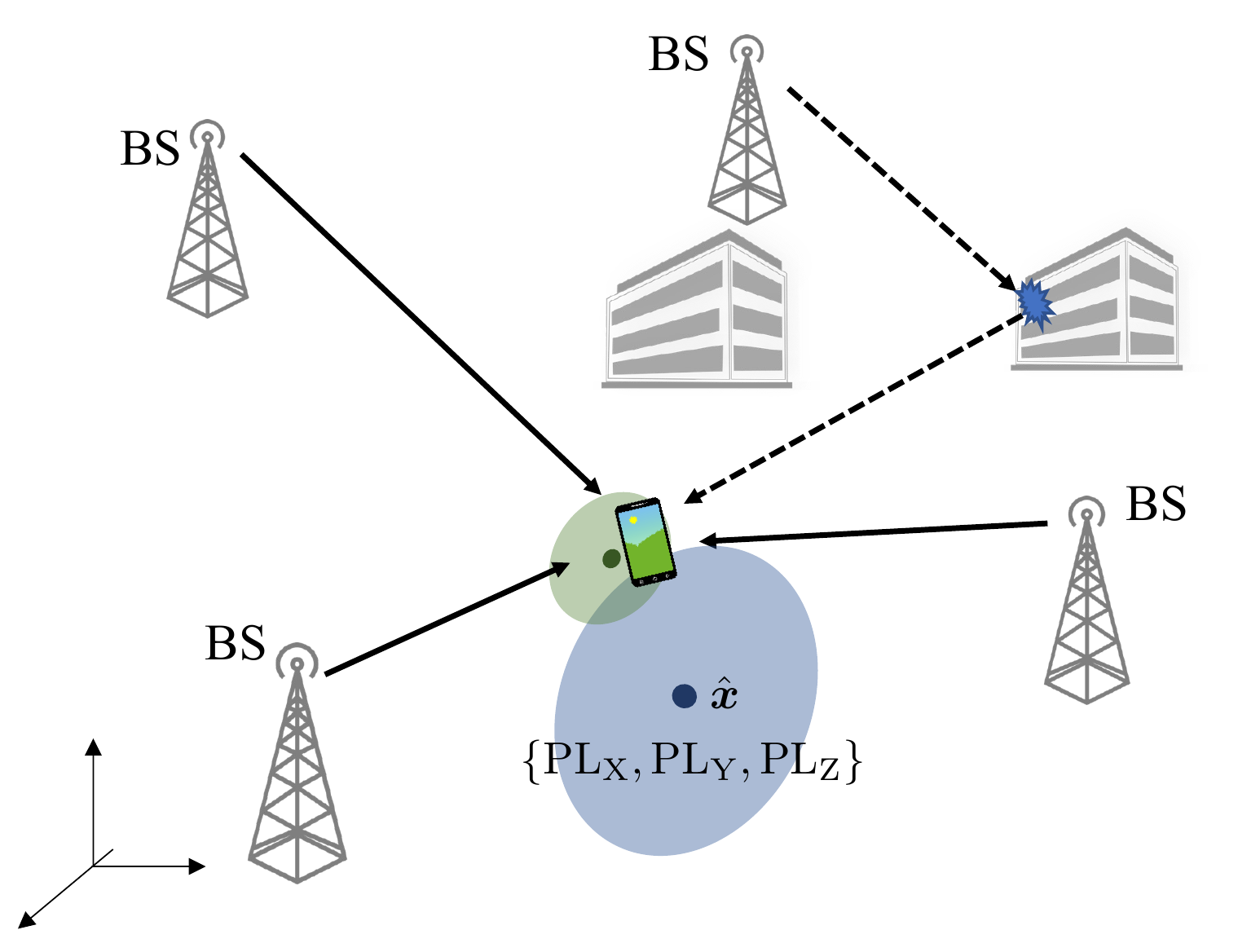} 
	\caption{Illustration of the cellular positioning with integrity monitoring problem. In the displayed scenario, one  \ac{BS} is in \ac{NLoS} condition due to blockage and causes a large bias to the pseudo-range measurement. When a \ac{RAIM} algorithm failed to exclude this faulty measurement, the computed position estimate $\hat{\boldsymbol{x}}$ and \acp{PL}  $\{ \mathrm{PL}_\mathrm{X}, \mathrm{PL}_\mathrm{Y}, \mathrm{PL}_\mathrm{Z}\}$ are represented by the blue ellipsoid are obtained; when it excludes the faulty measurement successfully, better results represented by the green ellipsoid are obtained. 
 } 
	\vspace{-5mm}
	\label{fig0}
\end{figure}

In this section, we describe the snapshot \ac{3D} positioning and integrity monitoring problem and its simplified \ac{1D} version. 

\subsection{The Snapshot Integrity Monitoring Problem}
We consider a scenario with $M$ time-synchronized \acp{BS} with known locations $\boldsymbol{x}_i \in \mathbb{R}^3$, $i \in \mathcal I \triangleq \{1,\ldots, M\}$, and a single mobile \ac{UE} with unknown position $\boldsymbol{x}=[x_\mathrm{X},x_\mathrm{Y},x_\mathrm{Z}]^\mathrm{T} \in \mathbb{R}^3$ and unknown clock bias $C \in \mathbb{R}$ (expressed in meters). Considering downlink/user-centric positioning, at each positioning epoch the \acp{BS} send coordinated \acp{PRS}, from which the \ac{UE} estimates $M$ \acp{ToA} from the \ac{LoS} paths, which are converted to pseudo-ranges as follows: 
\begin{align}
    y_i = \Vert \boldsymbol{x}_i - \boldsymbol{x} \Vert + C + b_i + n_i, 
    \label{eq:pseudorange}
\end{align}
for $i \in \mathcal I$, where $n_i$ is the independent measurement noise and $b_i$ is the range bias that accounts for any possible faults, such as synchronization errors or \ac{NLoS} biases. 
We further consider that an initial estimate of the \ac{UE} position is available, so that \eqref{eq:pseudorange} can be linearized \cite{koivisto2017joint, guvenc2012fundamental, zhu2009simple} to yield (after removing unnecessary terms) 
\begin{align} \label{eq:linearized-3D-model}
    y_i = \boldsymbol{a}_i^{\mathrm T}
\begin{bmatrix}
\boldsymbol{x}\\
C
\end{bmatrix}
 + b_i + n_i,
\end{align}
where $\boldsymbol{a}_i \in \mathbb{R}^{4 \times 1}$ is a known vector. 
The objective of integrity monitoring is to calculate a position estimate $\hat{\boldsymbol{x}}$ and a \ac{PL} for each dimension $n$ in 
$\{\mathrm{X}, \mathrm{Y}, \mathrm{Z} \}$  such that 
\begin{align}
     \mathrm{Pr}(|x_n-\hat{x}_n|> \mathrm{PL}_n) \leq \mathrm{TIR}_n,  
     \label{eq:3D-PL-computation}
\end{align}
where $\mathrm{Pr}(|x_n-\hat{x}_n|> \mathrm{PL}_n)$ is the actual \ac{IR} and $\mathrm{TIR}_n$ the \ac{TIR} for dimension $n$. A schematic drawing of the problem is shown in Fig.~\ref{fig0}. 

\subsection{Simplified Problem}
Inspired by the observation from \eqref{eq:3D-PL-computation}, that the PL is computed per dimension, we propose a simplified observation model, with the purpose of understanding the possible gains of Bayesian integrity monitoring over conventional frequentist approaches. In its most simple and nontrivial version, that is, a 1D positioning problem without clock bias and with Gaussian error, the 1D model analogy to \eqref{eq:linearized-3D-model} is 
\begin{align}\label{eq:1D_measurement_model}
    y_i = x + b_i + n_i, 
\end{align}
where $x$, $b_i$, and $n_i$, for $i \in \mathcal I$, are independent random variables. The measurement noise is modeled as $n_i \sim \mathcal{N}(n_i, 0, \sigma_{n,i}^2)$, where the notation $\mathcal{N}(t; m,\sigma^2)$ represents the Gaussian distribution for the random variable $t$ with mean $m$ and variance $\sigma^2$. 
A latent variable 
$\lambda_i \in \{ 0,1\}$ is adopted as an indicator of whether the measurement $y_i$ is faulty or not, which follows a Bernoulli \ac{PMF} given by $p_{\Lambda_i}(\lambda_i)=\theta_i^{\lambda_i}(1-\theta_i)^{(1-\lambda_i)}$, where $\theta_i$ is a known prior probability and $0< \theta_i \ll 1$ is assumed. 
When $\lambda_i= 0$, $y_i$ is free from fault and hence $b_i = 0$. When $\lambda_i= 1$, $b_i$ is modeled as a Gaussian random variable $\mathcal{N}(b_i; m_{b,i}, \sigma_{b,i}^2)$, whose variance $\sigma_{b,i}^2$ is considered considerably greater than $\sigma_{n,i}^2$. 
Consequently, the prior \ac{PDF} of $b_i$ is given by
\begin{align}
    p_{B_i}(b_i) 
    &=p_{B_i\mid \Lambda_i}(b_i\mid 0) \,p_{\Lambda_i}(0) + p_{B_i \mid \Lambda_i}(b_i\mid 1) \,p_{\Lambda_i}(1) \nonumber \\
    &= (1-\theta_i) \delta(b_i) + \theta_i \,\mathcal{N}(b_i; m_{b,i}, \sigma_{b,i}^2),
\end{align}
where $\delta(\cdot)$ denotes the Dirac delta distribution. 
The objective is then to find a position estimate $\hat{x}$ and a PL such that 
\begin{align}
    \mathrm{Pr}(|x -\hat{x}|> \mathrm{PL}) \leq \mathrm{TIR}.
\end{align}

\section{Baseline RAIM Solution}

For performance benchmarking purposes, we modify the hypothesis testing-based advanced RAIM algorithm \cite{blanch2015baseline} to the 1D problem described above. The original algorithm was developed in a multi-fault and multi-constellation GNSS setting. The algorithm has four components: (i) computation of the so-called all-in-view solution based on all $M$ measurements; (ii) identification of the fault modes that will be monitored, that is, which combination of \acp{BS} will be considered; (iii) detection of the faults and computation of PL if no faults were detected; (iv) exclusion of possibly faulty measurements if faults were detected and computation of PL. 

\subsection{All-in-View Solution}
We rewrite the measurement model \eqref{eq:1D_measurement_model} in vector form: 
\begin{align}
    \boldsymbol{y} = \boldsymbol{1}_M x + \boldsymbol{z}, 
\end{align}
where $\boldsymbol{y} = [y_1,\ldots, y_M]^\mathrm{T}$, $\boldsymbol{1}_M$ is a $M\times 1$ all-one vector and $\boldsymbol z = [ z_1, \ldots, z_M]^{\mathrm T}$ with elements $z_i = b_i + n_i$, $i = 1,\ldots, M$. 
When all measurements are free of fault, $\boldsymbol z \sim \mathcal N(\boldsymbol{0}, \boldsymbol{\Sigma})$, where $\boldsymbol{\Sigma}$ is the diagonal covariance matrix with diagonals $\boldsymbol{\Sigma}_{i,i} = \sigma_{n,i}^2$, $i=1,\ldots, M$, and the weighted least squares (WLS) estimate for $x$, which in this case also is the maximum likelihood (ML) estimate, is given by 
\begin{align}
    \hat{x}^{(0)} = \left( \boldsymbol 1_M^\mathrm{T} \boldsymbol{\Sigma}^{-1} \boldsymbol 1_M \right)^{-1} \boldsymbol 1_M^\mathrm{T} \boldsymbol{\Sigma}^{-1} \boldsymbol y = W_0 \boldsymbol{c}_0^\mathrm{T} \boldsymbol y, 
\end{align}
where $W_0 \triangleq  \left( \boldsymbol 1_M^\mathrm{T} \boldsymbol{\Sigma}^{-1} \boldsymbol 1_M \right)^{-1} = { \big( \sum_{i=1}^{M} {\sigma_{n,i}^{-2}} \big) }^{-1} $, and $\boldsymbol{c}_0^\mathrm{T} \triangleq \boldsymbol{1}_M^\mathrm{T} \boldsymbol{\Sigma}^{-1} =[ {\sigma_{n,1}^{-2} }, \ldots, {\sigma_{n,M}^{-2} } ]$.

\subsection{Fault Modes Identification}

First, the baseline \ac{RAIM} algorithm determines the set of \textit{fault modes} to be monitored, based on prior knowledge of the probability that each measurement is faulty. A fault mode is a specific subset of measurements that are simultaneously faulty, while the rest of the measurements are fault-free.  The baseline algorithm lists all fault modes that contain up to $M-2$ measurements\footnote{To be able to perform fault detection, as will be soon detailed, there should be at least $N_\mathrm{unknown} + 1$ measurements available, where $N_\mathrm{unknown}$ stands for the number of unknown position variables, which is $1$ in our \ac{1D} model.} and computes the corresponding probabilities of occurrence. Therefore, the total number of fault modes is given by $N_{\mathrm{FM}} =  \sum_{j=1}^{M-2} \binom{M}{j}$. 
For convenience, the fault-free case (i.e., the null hypothesis) is denoted as the fault mode $0$. 
We denote by $\mathcal{I}_k$ the set of faulty measurement indices contained in the fault mode $k\in \{0,1,\ldots, N_{\mathrm{FM}}\}$. In particular, $\mathcal{I}_0 =  \emptyset$.
The probability that this fault mode occurs is thus given by
\begin{align}
 p_{\mathrm{FM},k} = \prod_{i \in \mathcal{I}_k} \theta_i \prod_{i \in \mathcal{I}\setminus \mathcal{I}_k} (1 - \theta_i).    
\end{align}
Fault modes are sorted in decreasing order of their probability of occurrence. Namely, $p_{\mathrm{FM},k} \geq p_{\mathrm{FM},k+1}$, for $k<N_{\mathrm{FM}}$. Note that since $0< \theta_i \ll 1$ 
is commonly assumed, $p_{\mathrm{FM},0} > p_{\mathrm{FM},1}$ is considered true.

We note that to ensure the best possible performance, \emph{all} fault modes that can be monitored are listed in this baseline RAIM algorithm. In the original algorithm described in \cite{blanch2015baseline}, those fault modes that contain a large number of faulty measurements but with a very small probability of occurrence are left out of the monitoring, so that the computational complexity of the algorithm can be reduced.  We refer the interested readers to \cite{blanch2015baseline} for more details.

\subsection{Fault Detection}
For fault mode $k$, we define $W_k \triangleq  { \big( \sum_{i \in \mathcal I\setminus \mathcal I_k} {\sigma_{n,i}^{-2}} \big) }^{-1} $ and let $\boldsymbol{c}_k^\mathrm{T} $ to be the vector given by replacing elements of $\boldsymbol{c}_0^\mathrm{T}$ with indices in $\mathcal I_k$ by $0$. It can be easily verified that 
\begin{align}\label{eq:x_hat_k}
    \hat{x}^{(k)} = W_k \boldsymbol{c}_k^\mathrm{T} \boldsymbol{y} 
\end{align}
is the WLS estimate of $x$ using the remaining measures after excluding those contained in fault mode $k$ from $\boldsymbol{y}$.

The baseline RAIM algorithm determines whether the measurements contain faults by performing a list of \textit{\ac{SS} tests} for each fault mode.  
When all measurements are fault-free, it can be shown that the test statistic $\Delta \hat{x}^{(k)} \triangleq \hat{x}^{(0)} -\hat{x}^{(k)}$ is a Gaussian random variable with zero mean and variance given by 
\begin{align}
    \sigma_{ss}^{(k)2} \triangleq  \left( W_k \boldsymbol{c}_k^\mathrm{T} - W_0 \boldsymbol{c}_0^\mathrm{T} \right) \boldsymbol{\Sigma} \left( W_k \boldsymbol{c}_k^\mathrm{T} - W_0 \boldsymbol{c}_0^\mathrm{T} \right)^\mathrm{T}. 
\end{align}
To be able to identify the test thresholds, a false alarm probability $P_\mathrm{FA}$ is also required as input to the algorithm. The false alarm budget is evenly allocated to the $N_{\mathrm{FM}} $ fault modes that contain fault(s), leading to the following \ac{SS} test threshold for fault mode $k$, $k=1,\ldots, N_{\mathrm{FM}}$, 
\begin{align}\label{eq:T_ss_test}
    T_{k} = \sigma_{ss}^{(k)} \, Q^{-1} \left(\frac{P_\mathrm{FA}}{2 N_{\mathrm{FM}} }\right),  
\end{align}
where $Q^{-1}( \cdot)$ is the inverse of the $Q$ function, $Q(u) = \frac{1}{\sqrt{2\pi}}\int_u^{+\infty} e^{-\frac{t^2}{2}} \mathrm d t$. To be specific, $| \Delta \hat{x}^{(k)} |$ will be compared with $T_{k}$, and if $| \Delta \hat{x}^{(k)} | \leq T_k$, $\forall k = 1,\ldots, N_{\mathrm{FM}}$, all measurements are considered fault-free and the baseline RAIM algorithm outputs $\hat{x}^{(0)}$ as the position estimate. 
The choice of $P_\mathrm{FA}$, therefore, affects the continuity and availability performance of the positioning system: the larger the value, the more likely the algorithm is to warn upon a potential fault and will put more computational effort into fault exclusion attempts.

A \ac{PL} is computed by solving the following equation: 
\begin{align}\label{eq:PL_ARAIM}
    2  Q \left( \frac{ \mathrm{PL} }{\sigma^{(0)}} \right) 
    +  \sum_{k=1}^{N_\mathrm{FM}} p_{\mathrm{FM},k} \, Q\left( \frac{\mathrm{PL} - T_{k}}{\sigma^{(k)} } \right) = \mathrm{TIR}, 
\end{align}
where $\sigma^{(k)} \triangleq W_k \sqrt{ \boldsymbol{c}_k^\mathrm{T}  \boldsymbol{\Sigma} \boldsymbol{c}_k }$, and $Q\left( \frac{\mathrm{PL} - T_{k}}{\sigma^{(k)} } \right)$ is an upper bound of the actual \ac{IR} when fault mode $k$ occurs but the \ac{SS} test passes, i.e., $\mathrm{Pr}(|x -\hat{x}^{(0)}|> \mathrm{PL}, | \hat{x}^{(0)} -\hat{x}^{(k)} | \leq T_k)$. The formulation process of the above equation can be found in \cite[Appendix H]{blanch2015baseline}, and a solution of \eqref{eq:PL_ARAIM} is found using the bisection search method detailed in  \cite[Appendix B]{blanch2015baseline}.

\subsection{Fault Exclusion}

When any \ac{SS} tests fail, that is, $| \Delta \hat{x}^{(k)} | > T_k$ for some $k$, the baseline RAIM algorithm will try to exclude faulty measurements that cause failure. Starting from $k=1$, 
the algorithm performs the \textit{complete} process of fault modes identification and fault detection on the \textit{new} problem formed after removing the measurements contained in fault mode $k$. 
The remaining number of measurements in this new problem is $M-|\mathcal I_k|$. If the \ac{SS} tests pass successfully, the fault exclusion process is terminated and a PL can be computed in the same way as described above (but using the newly obtained values for $p_{\mathrm{FM},k}$, $T_k$, $N_\mathrm{FM}$, etc.). 
Otherwise, the algorithm continues to check the next fault mode ($k+1$). If the \ac{SS} tests fail for all fault modes, then the fault exclusion attempt is considered failed. In this case, the algorithm terminates without being able to compute a PL and should claim that the position estimate should not be trusted.

\section{Proposed Bayesian Methods}

From a Bayesian perspective, we treat $x$ and $\lambda_i$ in \eqref{eq:1D_measurement_model} as random variables, whose prior probability distributions, $p_X(x)$
and $ p_{\Lambda_i}(\lambda_i)$, $\forall i\in\mathcal{I}$, are known.\footnote{Note that $p(x)$ and $ p(\lambda_i)$ may be passed on from the previous epoch(s), where $p(x)$ requires a \ac{UE} mobility model. In the absence of prior information, $p(x)$ and $ p(\lambda_i)$ can be set to uniform distributions. } 
For each epoch of the integrity monitoring problem, we aim to determine the marginal posterior distribution of each random variable, i.e., $ p_{X|\boldsymbol Y}(x | \boldsymbol{y})$ and $ p_{\Lambda_i|\boldsymbol Y}(\lambda_i | \boldsymbol{y})$, $i\in\mathcal{I}$, from the prior information, the measurements $\boldsymbol{y}$, and the corresponding likelihood function induced by \eqref{eq:1D_measurement_model}. Based on the posterior distributions, different methods can be selected to compute the position estimate $\hat{x}$ and PL to meet the \ac{TIR} requirement.

\subsection{Message Passing Algorithm}
\label{sec:3a}

\begin{figure}[!t]
	\centering
\includegraphics[width=\linewidth]{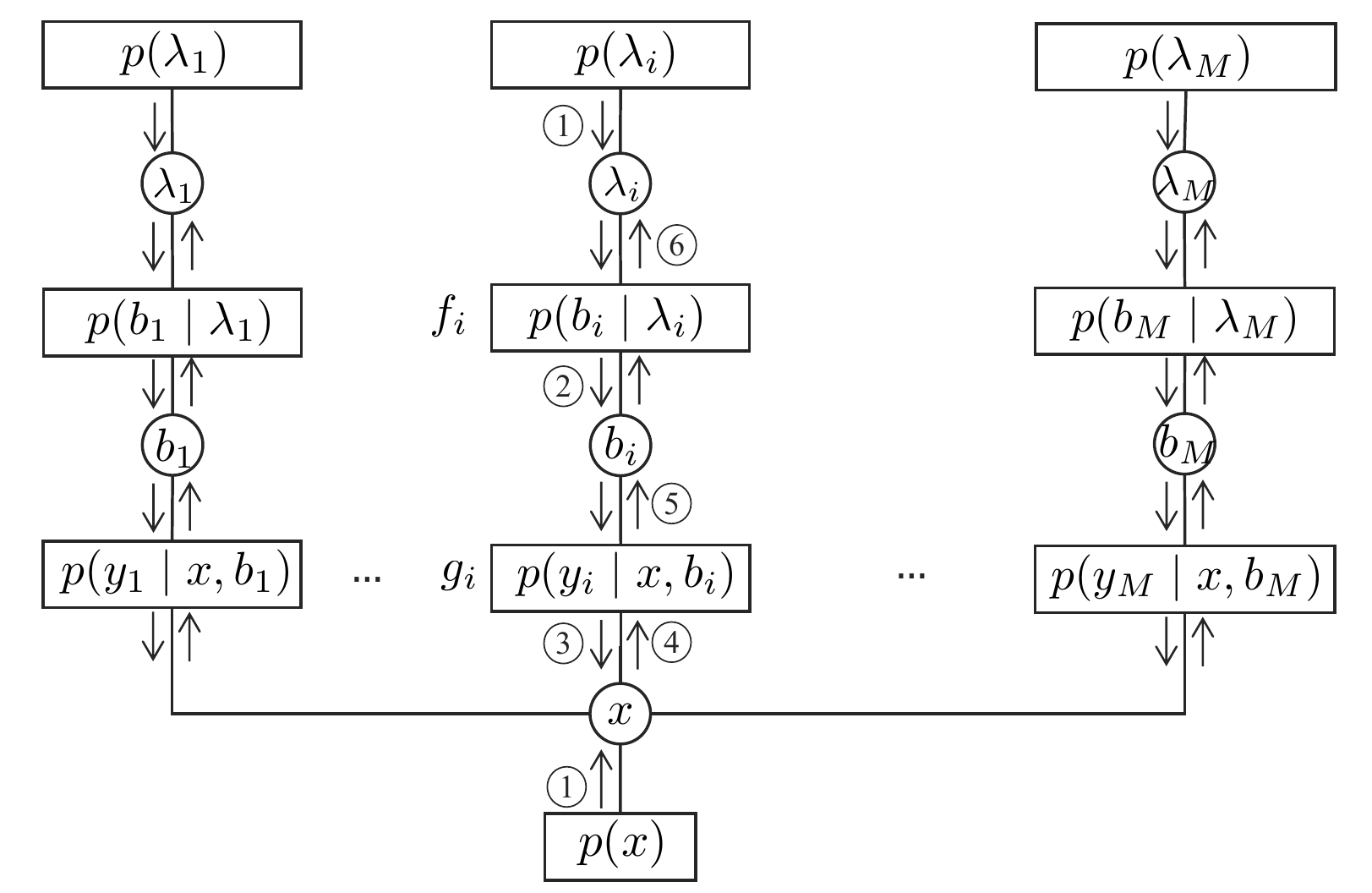} 
	\caption{Factor graph of the Bayesian RAIM problem, corresponding to the factorization \eqref{eq:joint_posterior}. The order of message computation and passing is given by the numbers in circles (shown only on the $i$th branch, but is the same for all branches), while the arrows indicate the passing direction.} 
	\label{fig1}
\end{figure}

Based on the assumptions made in Section \ref{sec:2}, 
the joint posterior probability of $x$, $\boldsymbol{b} \triangleq [b_1, \ldots, b_M]^{\mathrm{T}}$ and $\boldsymbol{\lambda} \triangleq [\lambda_1,\ldots,\lambda_M]^{\mathrm{T}}$ can be factorized as 
\begin{align} 
p(x,\boldsymbol{b},\boldsymbol{\lambda} \mid \boldsymbol{y}) 
&\propto p(\boldsymbol{y} \mid x,\boldsymbol{b},\boldsymbol{\lambda}) p(x,\boldsymbol{b},\boldsymbol{\lambda}) \nonumber \\
& = p(x) \prod_{i=1}^{M} p(y_i \mid x,b_i) p(b_i \mid \lambda_i) p(\lambda_i).  \label{eq:joint_posterior}
\end{align}
For clarity, the subscripts of the probability distributions are omitted in \eqref{eq:joint_posterior} and all subsequent expressions and the variables to which they belong should be clear from the context. 

A cycle-free factor graph representation of \eqref{eq:joint_posterior} is shown in Fig.~\ref{fig1}, where each term of \eqref{eq:joint_posterior} is represented by a factor node in a rectangle and connected to the variable nodes that appear in parentheses, represented by circles in the graph.  The leaf (factor) nodes are all prior distribution functions. The remaining two categories of factor nodes are the conditional density functions $p(b_i | \lambda_i)$ and $p(y_i | x, b_i)$, which are denoted by $f_i$ and $g_i$ respectively for simplicity. 
Since the factor graph is cycle-free, a simple message-passing schedule is applied following the sum-product algorithm \cite{kschischang2001factor}. 

The following notations and operations are involved.  
$\mathrm{GM}(t, L)$ stands for a \ac{GM} distribution for $t$ with $L$ components: $\sum_{l=1}^{L} w_l \mathcal{N}(t; m_l,\sigma_l^2)$ with $\sum_{l=1}^{L} w_{l} =1$.  
Many messages passed on the factor graph are of this form, and their weights $\{w_l\}$, means $\{m_l\}$, and standard deviations $\{\sigma_l\}$ are what actually needs to be sent. 
Given $\mathrm{GM}(t, L_1) = \sum_{l_1=1}^{L_1} w_{l_1} \mathcal{N}(t; m_{l_1},\sigma_{l_1}^2)$ and $\mathrm{GM}(t, L_2) = \sum_{l_2=1}^{L_2} w_{l_2} \mathcal{N}(t; m_{l_2},\sigma_{l_2}^2)$,
their product is given by
\begin{align}\label{eq:GM_c1}
   \sum_{l_1=1}^{L_1}  \sum_{l_2=1}^{L_2} w_{l_1}  w_{l_2} s_{l_1 l_2} \, 
   \mathcal{N}(t;m_{l_1 l_2},\sigma_{l_1 l_2}^2) \propto \mathrm{GM}(t,L_1L_2),
\end{align}
where $s_{l_1 l_2}$, $m_{l_1 l_2}$ and $\sigma_{l_1 l_2}$ are obtained by solving the following simple equations: 
\begin{align}\label{eq:GM_c2}
    \begin{cases}
        {1}/{\sigma_{l_1 l_2}^2} = {1}/{\sigma_{l_1}^2} + {1}/{\sigma_{l_2}^2},
 \\
    {m_{l_1 l_2}}/{\sigma_{l_1 l_2}^2} = {m_{l_1}}/{\sigma_{l_1}^2}  + {m_{l_2}}/{\sigma_{l_2}^2}, 
\\
s_{l_1 l_2}  = \mathcal{N} (m_{l_1}; m_{l_2}, \sigma_{l_1}^2 + \sigma_{l_2}^2). 
    \end{cases}
\end{align}
A limited number of arithmetic operations are needed to solve these equations. Therefore, the computational complexity associated with this product is given by $\mathcal{O}(L_1 L_2)$.

The message-passing schedule is described below for the $i$th branch. It proceeds in parallel along all $M$ branches. 
\begin{enumerate}
    \item[\circled{1}]  The prior $p(\lambda_i)$ is sent from the leaf node to variable nodes $\lambda_i$, and then directly to $f_i$. 
    Meanwhile, $p(x)$ is sent to the variable node $x$. 
    \item[\circled{2}]  
    Factor node $f_i$ sends the following message to variable node $b_i$: 
    \begin{align}
    \mu_{f_i \rightarrow b_i} (b_i) 
    &= \sum_{ \lambda_i } p(b_i \mid \lambda_i) p(\lambda_i) .
    \end{align}
    which is a \ac{GM} distribution of $b_i$, given by $ \mu_{f_i \rightarrow b_i} (b_i) = (1-\theta_i) \, \mathcal{N}(b_i;0,0) + \theta_i \, \mathcal{N}(b_i;m_{b,i},\sigma_{b,i}^2)$. 
    This message will be passed directly to factor node $g_i$. 
    \item[\circled{3}] Factor node $g_i$ sends the following message to variable node $x$: 
    \begin{align}
     \mu_{g_i \rightarrow x} (x) 
     = \int p(y_i \mid x, b_i) \, \mu_{f_i \rightarrow b_i} (b_i) \, \mathrm{d} b_i .
    \end{align} 
    Since $p(y_i | x, b_i) = \mathcal N(b_i; y_i-x,\sigma_{n,i}^2)$, based on \eqref{eq:GM_c1} and \eqref{eq:GM_c2}, the integration results in a $\mathrm{GM}(x, 2)$ density:
    \begin{align}
      (1-\theta_i) \, \mathcal{N}(x; y_i, \sigma_{n,i}^2) + \theta_i \, \mathcal{N}(x; y_i-m_{b,i},\sigma_{n,i}^2+\sigma_{b,i}^2).
    \end{align} 
    \item[\circled{4}] Variable node $x$ sends the following message back to $g_i$: 
    \begin{align}
    \mu_{x \rightarrow g_i} (x) &= p(x) \prod_{j \in \mathcal M \setminus i}  \mu_{g_j \rightarrow x} (x) .
    \end{align}
    Under the assumption of Gaussian or uniform distribution of $p(x)$, $\mu_{x \rightarrow g_i} (x) \propto \mathrm{GM}(x, 2^{M-1})$. 
    \item[\circled{5}] Factor node $g_i$ sends the following message back to $b_i$: 
    \begin{align}
    \mu_{g_i \rightarrow b_i} (b_i) &= \int p(y_i \mid x, b_i) \, \mu_{x \rightarrow g_i} (x) \, \mathrm{d} x ,
    \end{align}
    which is then passed on to factor node $f_i$ directly. Since $p(y_i | x, b_i)$ can also be regarded as Gaussian \ac{PDF} of $x$, i.e., $\mathcal N(x; y_i-b_i,\sigma_{n,i}^2)$, following the same procedure as in step \circled{3}, it is clear that $\mu_{g_i \rightarrow b_i} (b_i) \propto \mathrm{GM}(b_i, 2^{M-1}) $ and the parameters are easy to compute. 
    \item[\circled{6}] Finally, the factor node $f_i$ sends a message back to $\lambda_i$, which is given by 
    \begin{align}
    \mu_{f_i \rightarrow \lambda_i} (\lambda_i) &= \int p(b_i \mid  \lambda_i) \, \mu_{g_i \rightarrow b_i} (b_i) \, \mathrm{d} b_i.  
    \end{align}
    The message is computed separately for $\lambda_i=0$ and $\lambda_i=1$. In particular, $\mu_{f_i \rightarrow \lambda_i} (\lambda_i = 0) = \mu_{g_i \rightarrow b_i} (b_i =0)$, and the computation of $\mu_{f_i \rightarrow \lambda_i} (\lambda_i = 1)$ requires $2^{M-1}$ terms of product of two Gaussian in the integrand. 
\end{enumerate}

The message-passing process terminates after the above six steps. The marginal posteriors can be calculated by multiplying all the incoming messages at the variable nodes $x$ and $\{\lambda_i\}$. In particular, 
\begin{align}\label{eq:x_postPDF1}
    p(x\mid \boldsymbol{y}) \propto p(x) \prod_{i \in \mathcal M}  \mu_{g_i \rightarrow x} (x), 
\end{align}
which leads to a GM distribution with $2^M$ terms, and 
\begin{align}\label{eq:lambda_post}
    p(\lambda_i \mid \boldsymbol{y}) = \begin{cases}
        1 - \theta_i'  \propto (1\!-\!\theta_i)\cdot  \mu_{f_i \rightarrow \lambda_i} (\lambda_i =0) , & \text{if } \lambda_i =0, \\ 
        \theta_i' \hspace{6mm}\propto \theta_i \cdot \mu_{f_i \rightarrow \lambda_i} (\lambda_i =1) , & \text{if } \lambda_i = 1,
    \end{cases}
\end{align}
where $\theta_i'$ can be obtained after normalization.

\subsection{Fault Exclusion}
\label{sec:3b}

In theory, all the information about the user position contained in the measurements and priors has been collected in $p(x | \boldsymbol{y})$ given by \eqref{eq:x_postPDF1}, and a position estimate and a PL can be computed directly based on it. However, the bias term in a faulty measurement will introduce a large uncertainty in the posterior, which will lead to a large PL. To counteract this effect, fault exclusion can be performed based on the marginal posterior \acp{PMF} of the indicators given by \eqref{eq:lambda_post}, before computing the marginal posterior \ac{PDF} of $x$ using the remaining measurements. 
To be specific, a threshold $\theta_{T}$ is selected, and if $\theta_i' > \theta_{T}$, the measurement $y_i$ is considered faulty, the corresponding branch will be pruned from the factor graph and the message $\mu_{g_i \rightarrow x} (x)$ passed to variable node $x$ along it will be discarded. No other additional message computation/passing is required. 

Denoting the set of indices of excluded measurements is by $\mathcal{I}_{\mathrm{F}}$, the marginal posterior of $x$ after fault exclusion is computed by
\begin{align}\label{eq:x_postPDF2}
    p_{\mathrm{FE}}(x\mid \boldsymbol{y}) \propto p(x) \prod_{ i\in \mathcal{I} \setminus \mathcal{I}_{\mathrm{F}} } \mu_{g_i \rightarrow x} (x), 
\end{align}
which leads to a GM posterior distribution of $ 2^{M-|\mathcal{I}_{\mathrm{F}}|}$ terms.

\subsection{Position Estimation and PL Computation}

With or without fault exclusion, the marginal posterior of $x$ can anyway be written in the following form: 
\begin{align}\label{eq:x_GM}
    p_{X}^{\mathrm{pos}}(x) = \sum_{l=1}^{L} w_{x,l} \, \mathcal{N}(x; m_{x,l},\sigma_{x,l}^2), 
\end{align}
where $\{w_{x,l}, l =1,\ldots,L\}$ is sorted in decreasing order and $L$ is either given by $ 2^{M}$ following \eqref{eq:x_postPDF1}, or by $2^{M -|\mathcal{I}_{\mathrm{F}}|}$ following \eqref{eq:x_postPDF2}. 
For position estimation, the weighted mean (WM) method is adopted, so that $\hat{x} = \sum_{l=1}^{L} w_{x,l} \, m_{x,l}$.

Based on \eqref{eq:x_GM}, the \ac{IR} associated with a position estimate $\hat x$ and a said \ac{PL} can be formulated as 
\begin{align}
    \mathrm{IR}
    &= \mathrm{Pr}( x<  \hat{x} - \mathrm{PL}) +  \mathrm{Pr}( x > \hat{x} + \mathrm{PL}) \nonumber \\
    &= \sum_{l=1}^{L} w_{x,l} \Big[ \Phi_{\mathcal{N},l}(\hat{x} - \mathrm{PL} ) + 1 - \Phi_{\mathcal{N},l}(\hat{x} + \mathrm{PL} ) \Big], 
    \label{eq:PL_FG1}
\end{align}
where $\Phi_{\mathcal{N},l}(\cdot) $ stands for the \ac{CDF} of the $l$th Gaussian term in \eqref{eq:x_GM}. 
Using the notation of $Q$ function as in \eqref{eq:PL_ARAIM}, the goal is to find the smallest value for $\mathrm{PL}$ such that the following inequality holds: 
\begin{align}
    &\sum_{l=1}^{L} w_{x,l} \left[ 1 - Q \Big( \frac{\hat{x}\! -\! \mathrm{PL}\! -\! m_{x,l} }{\sigma_{x,l}}\Big) +  Q \Big( \frac{\hat{x}\! + \!\mathrm{PL}\! -\! m_{x,l} }{\sigma_{x,l}}\Big)\right] \nonumber\\
    &\leq \mathrm{TIR}. 
    \label{eq:PL_FG2}
\end{align}
For this, a bisection search is again employed by the Bayesian RAIM algorithms. 

Depending on whether to perform fault exclusion or not, two variations of the Bayesian RAIM algorithm have resulted under the same framework (labeled using FE or NFE in the numerical study). Their performance will be compared with the baseline RAIM algorithm in the following section. 
We also remark that other position estimation methods, e.g., maximum a posterior (MAP) estimation, can also be adopted, which can be expected to have an impact on the computed \ac{PL} according to \eqref{eq:PL_FG2}. The comparison of different position estimation methods will be left for future work.

\subsection{Computational Complexity Comparison}

For all other steps except PL computation, the Bayesian RAIM algorithm has a fixed complexity scaling of $\mathcal{O} ( 3 M \times  2^{M-1})$, mainly due to the computation of \eqref{eq:x_postPDF1} via message passing. The baseline RAIM have a variable complexity (as the number of \ac{SS} test varies), between $\mathcal{O} \left( N_{\mathrm{FM} } \right)$ (when all \ac{SS} tests pass in the fault detection step) and approximately $\mathcal{O} \left( N_{\mathrm{FM} }^2 \right)$ (when every fault mode is examined in the fault exclusion step), where $N_{\mathrm{FM} } = \sum_{j=1}^{M-2} \binom{M}{j} =  2^M - M -2$. 

Both methods require a bisection search to compute the \ac{PL}, which is an iterative algorithm. For a given number of iterations, say $N_\mathrm{it}>0$, the complexity of the PL computation in Bayesian RAIM is $\mathcal{O}(N_\mathrm{it}2L)$, where  $L = 2^{M}$ without fault exclusion and $L=2^{M -|\mathcal{I}_{\mathrm{F}}|}$ with fault exclusion. For baseline RAIM, this complexity is $\mathcal{O}(N_\mathrm{it}N_{\mathrm{FM}})$.

\begin{remark}
In cellular positioning, there are usually few connected \acp{BS} ($M \leq 10$). Therefore, the computational complexities of the baseline and proposed Bayesian RAIM algorithms can be said in a similar order for the \ac{1D} positioning problem. 
We note that in \ac{GNSS}, the number of satellites visible to the \ac{UE} can be many more, which could entail high computational complexities of the Bayesian algorithms. 
\end{remark}

\section{Numerical Study}
\label{sec:4}
\subsection{Scenario}
For numerical study, scenarios with $M= 5$ or $8$ \acp{BS} and one \ac{UE} are considered. Without loss of generality, the actual position of the UE is always at $x=0$. All \acp{BS} have the same measurement noise level $\sigma_{n,i}$ (thus denoted simply using $\sigma_{n}$ in the following discussion), which increases from $1$ to $9$ meters in steps of $2$. Moreover, $\forall i = 1,\ldots, M$, $\theta_i = 0.05$, $\sigma_{b,i} =50$ meters, and $m_{b,i} \sim \mathrm{uni}[-50,50]$ is randomly chosen and fixed during each set (corresponding to a certain pair of values for $M$ and $\sigma_{n}$) of simulations. The \ac{TIR} is set to $10^{-3}$, and at least $5\times 10^6$ independent realizations are simulated in each set of simulations. 
For the Bayesian RAIM algorithm with fault exclusion, the threshold $\theta_T = 0.5$ is adopted. For the baseline RAIM algorithm, $P_\mathrm{FA} = 5 \times 10^{-2}$ is selected. 

\subsection{Results and Discussion}

\begin{figure}[!t]
	\centering
	\subfigure[Bayesian RAIM with fault exclusion]{ \includegraphics [width= .84\linewidth,trim= 65 210 65 235, clip]{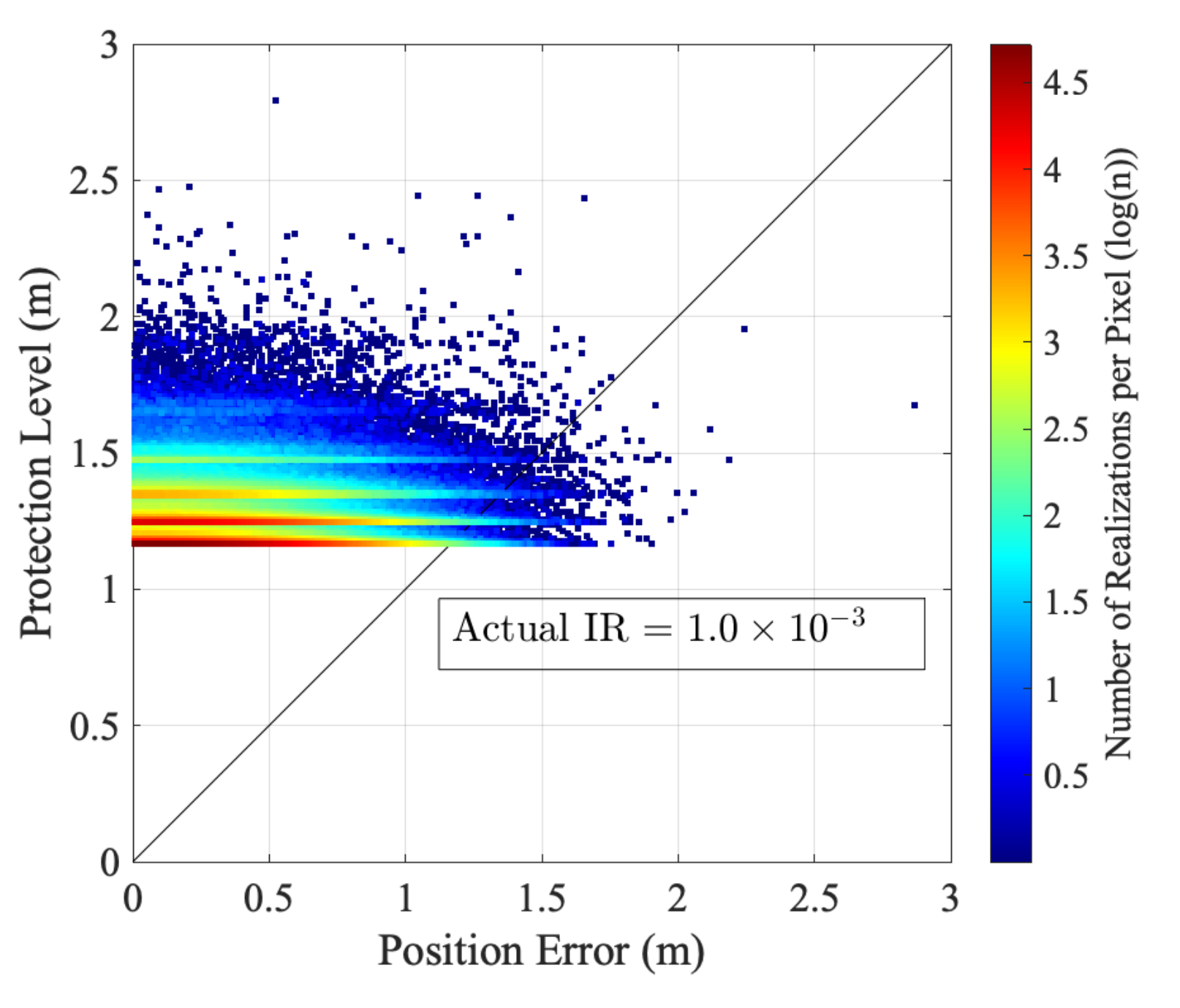}  }\vspace{-.4em} 
	\subfigure[Baseline RAIM]{\includegraphics [width= .84\linewidth,trim= 65 210 65 235, clip]{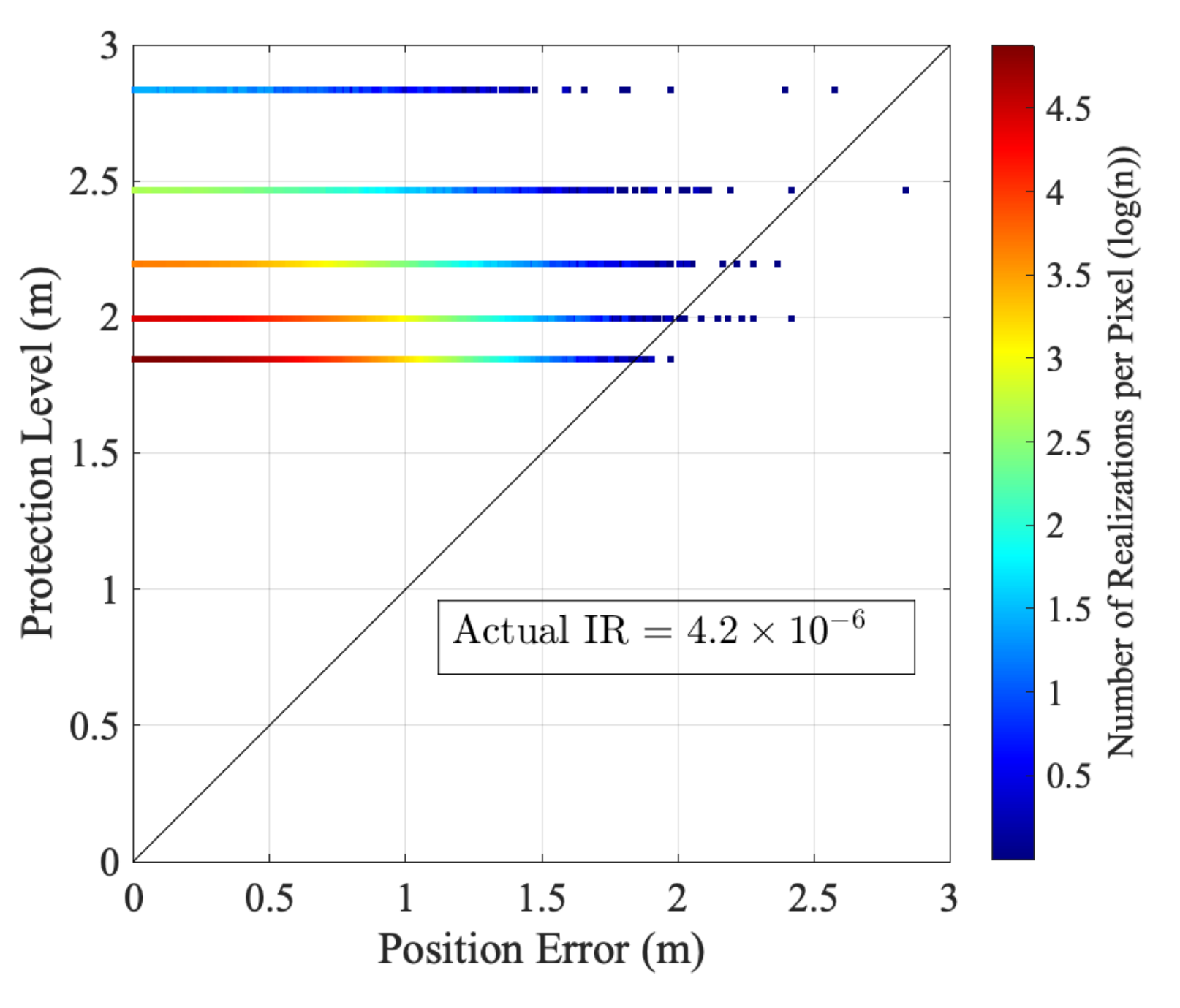} }
	\caption{Stanford diagrams of the proposed Bayesian RAIM algorithm with fault exclusion in (a) and the baseline RAIM algorithm in (b). Plotted for $M=8$ and $\sigma_{n} =1$ over $5\times 10^6$ realizations, and with a $1$ cm $\times 1$ cm pixel size. A dot that appears on the bottom right side of the diagonal line represents an integrity failure (i.e., $|x - \hat{x}| > \mathrm{PL}$); the simulated \ac{IR} of the Bayesian RAIM (a) is much tighter to the desired bound $\leq$ \ac{TIR} ($10^{-3}$). } 
	\label{fig:stanford}
 \vspace{-5mm}
\end{figure}

In each set of simulations, the simulated \ac{IR} is obtained, that is, the ratio of realizations when the actual position error $|x - \hat{x}|$ exceeds the computed \acp{PL} among all realizations. It is found that the simulated \acp{IR} resultant from the Bayesian RAIM algorithms, with or without fault exclusion, all converged to the \ac{TIR} after running enough realizations. This in turn proves that the posterior probabilities computed via the factor graph are exact. On the other hand, the simulated \ac{IR} results of the baseline RAIM algorithm are in the order of $10^{-6}$, much lower than the \ac{TIR}. 
For visualization,  the results given by the Bayesian RAIM algorithm (with fault exclusion) and the baseline RAIM algorithm (also with fault exclusion) in the setting with $M=8$ \acp{BS} and $\sigma_{n}=1$ meter over $5\times 10^6$ realizations are presented in the form of Stanford diagram \cite{whiton2022cellular}
in Fig.~\ref{fig:stanford}. 
As can be seen, the proposed Bayesian RAIM provides much tighter \acp{PL} while meeting the \ac{TIR} requirement, compared to the baseline RAIM. The latter is conservative in \ac{PL} computation, so the computed \acp{PL} are larger and the actual \ac{IR} is significantly smaller than the \ac{TIR}.
One can also see that the \ac{PL} results the baseline RAIM algorithms returns are discrete. This is because, unlike the Bayesian methods, the actual measurements are not utilized in the \ac{PL} computation, as can be seen from \eqref{eq:PL_ARAIM}. In particular, in \eqref{eq:PL_ARAIM}, $N_{\mathrm{FM}}$ is determined by the number of measurements excluded, and $p_{\mathrm{FM},k}$, $T_k$ and $\sigma^{(k)}$ can only take a limited number of values, since the measurement models are assumed to be the same for all \acp{BS}. 

\begin{figure}[!t]
	\centering
 \subfigure[$\sigma_{n} = 1$ meter]{ \includegraphics [width= .9\linewidth,trim= 40 240 50 250, clip]{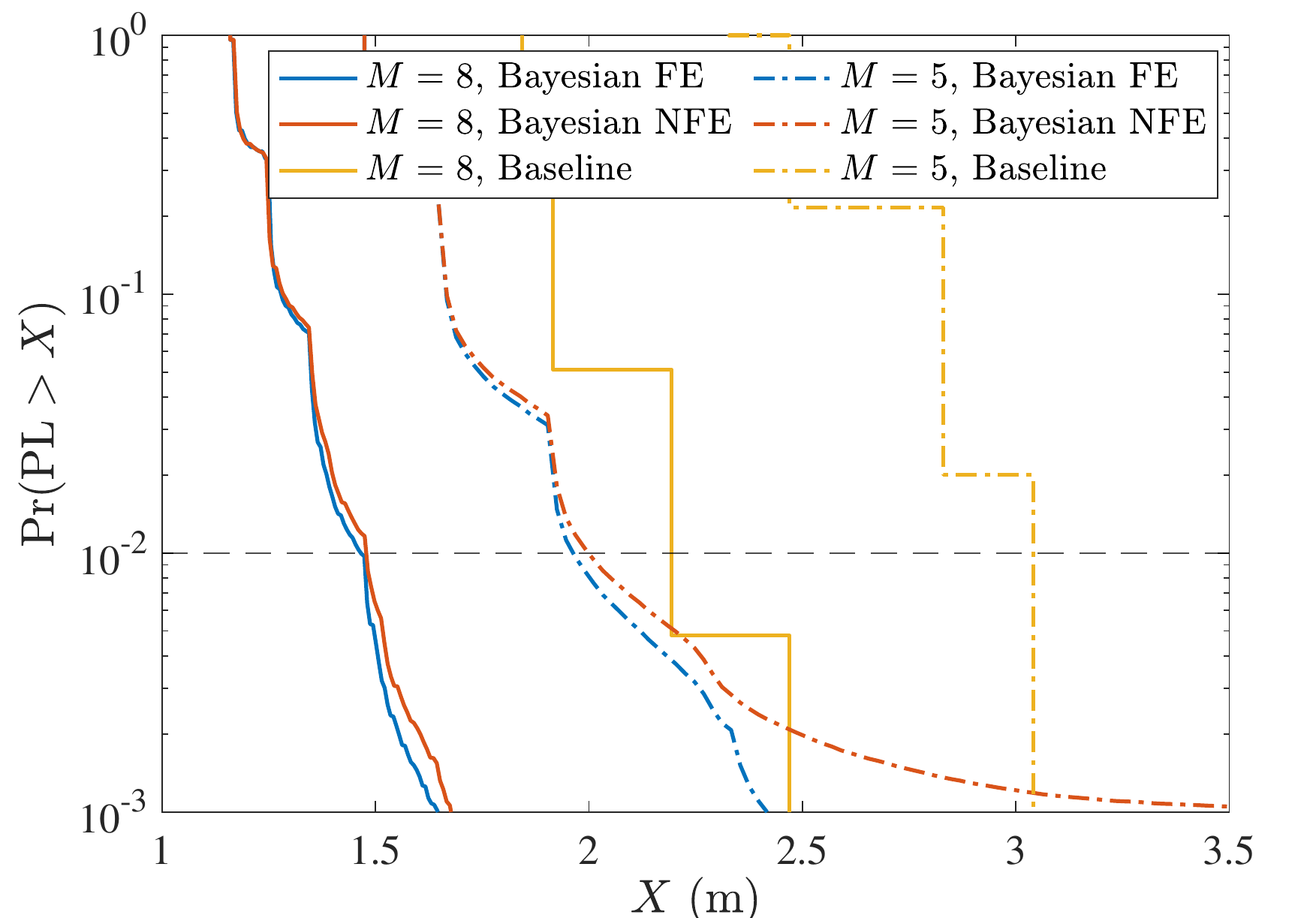}   }
 \subfigure[$\sigma_{n} = 9$ meters]{ \includegraphics [width= .9\linewidth,trim= 40 240 50 250, clip]{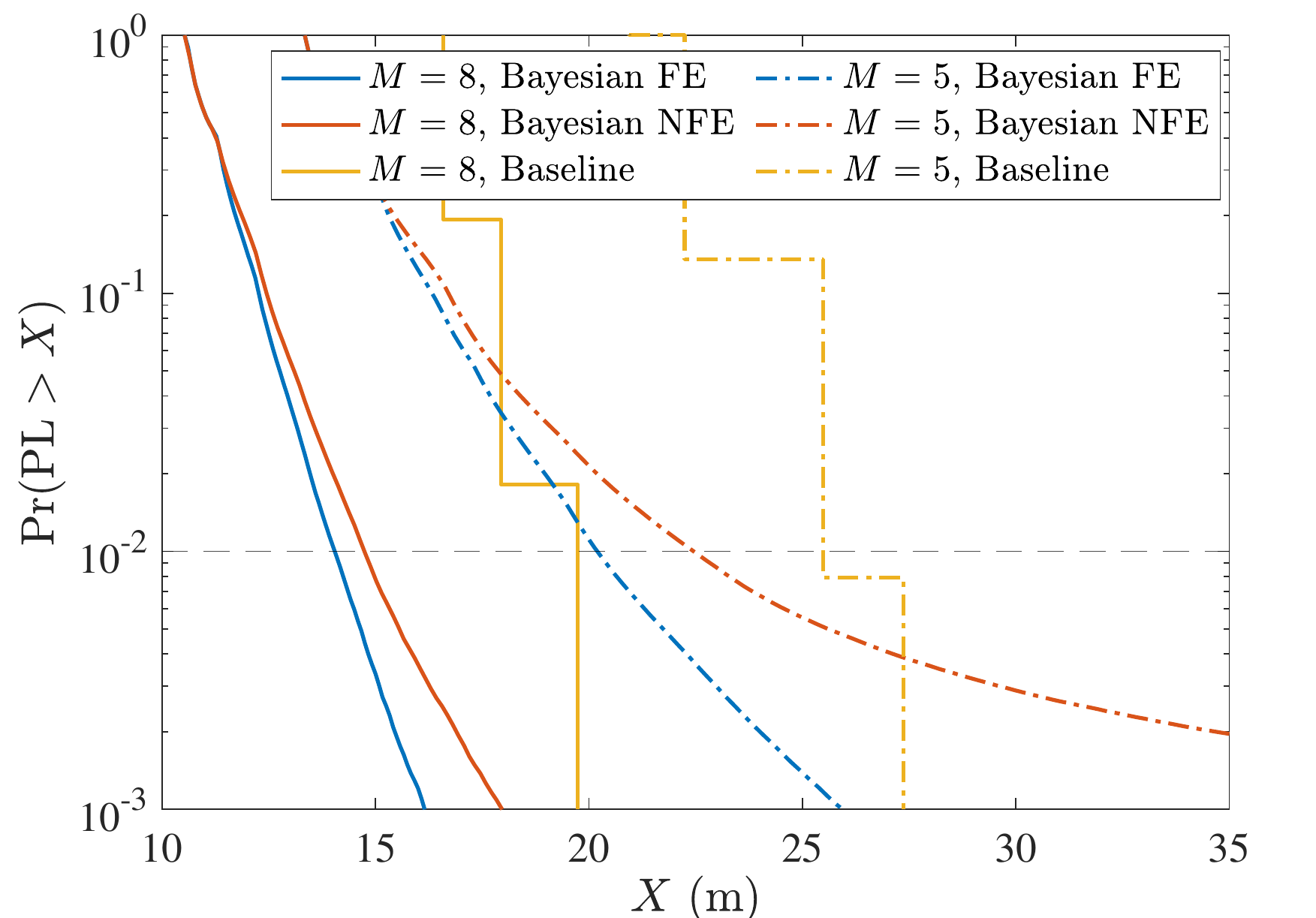}   }
	\caption{CCDF curves of PLs under two values of the measurement noise level.}  
	\label{fig:PL}
\end{figure}

\begin{figure}[!t]
	\centering
	\includegraphics [width= .9\linewidth,trim= 40 240 50 250, clip]{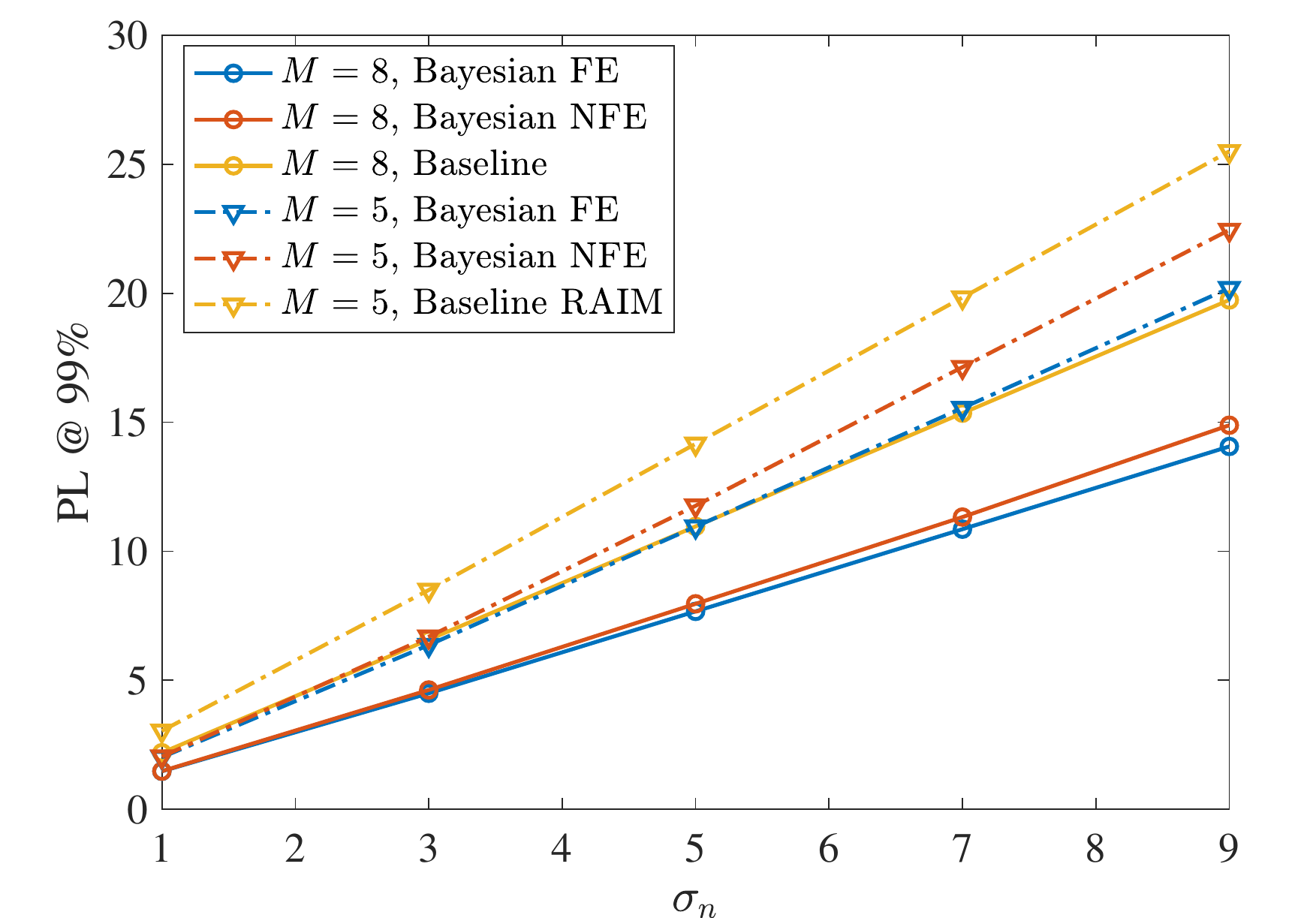} 
	\caption{PL values at $99\%$ percentile as a function of the measurement noise level.}  
	\label{fig:PL99}
\end{figure}

For each set of simulations, empirical \ac{CCDF} curves of the computed \acp{PL} are obtained. Four sets of the \ac{CCDF} curves are presented in Fig.~\ref{fig:PL}, from which the significant performance improvement of the proposed Bayesian RAIM algorithms over the baseline RAIM algorithm in obtaining tighter \acp{PL} can be clearly seen. 
In addition, the importance of fault exclusion before \ac{PL} computation is revealed. Comparing the \ac{PL} results given by the Bayesian RAIM algorithms with and without fault exclusion, we see that the large uncertainty introduced by the potentially faulty measurements in the posteriors leads to larger \ac{PL} results (in a statistical sense), and the gap increases when $\sigma_{n}$ increases or/and when $M$ decreases, because the uncertainty can be reduced by providing more fault-free measurements or by reducing the measurement noises. 
To better illustrate this effect, in Fig.~\ref{fig:PL99}, the \ac{PL} values at $99\%$ percentile (given by the intersections of the horizontal line at $10^{-2}$ with the \ac{CCDF} curves) obtained under all sets of simulations are shown. Another interesting observation from Fig.~\ref{fig:PL99} is that the $99\%$ percentile \ac{PL} results given by any evaluated RAIM algorithms seem to increase linearly with $\sigma_{n}$.

\section{Conclusion}

A Bayesian RAIM algorithm has been developed for snapshot-type positioning problems under a \ac{1D} linear Gaussian setting, as a methodological validation. Position estimation, multi-fault detection and exclusion, and \ac{PL} computation are performed based on the exact posterior distributions computed via message passing along a factor graph. It has been shown using Monte-Carlo simulation that the proposed algorithm provides tight \acp{PL} while satisfying the given \ac{TIR} requirement, leading to a significant performance improvement over the baseline advanced RAIM algorithm. 
Encouraged by the result, the algorithm will be extended to the \ac{ToA}-based \ac{3D} positioning problem under the same framework.

\bibliographystyle{IEEEtran}
\bibliography{integrity.bib}

\end{document}